\documentstyle[aps,prl,epsfig]{revtex}
\begin{document}
\wideabs{
\title{Role of twin boundaries on the vortex dynamics in YBa$_{2}$Cu$_{3}$O$_{7}$}
\author{H. Pastoriza\cite{conicet}, S. Candia\cite{cuyo} and G. Nieva\cite{conicet}}
\address{Comisi\'on Nacional de Energ\'{\i}a
At\'omica, Centro At\'omico Bariloche and Instituto Balseiro, \\8400 San Carlos de Bariloche, Argentina}
\preprint{V3.0}
\draft
\maketitle
\begin{abstract}

By means of a novel technique of rotating the applied current we have
directly measured the influence of twin boundaries on the vortex motion in a
YBa$_{2}$Cu$_{3}$O$_{7}$ single crystal.  The results indicate that the
effect of twin planes on the vortex dynamics starts to develop below a certain
temperature, being responsible for an anisotropic viscosity in the vortex
liquid state and a guided motion in the solid state.

\end{abstract}
}
\narrowtext

The study of vortex matter in superconductors has great interest from
several points of view. Technologically it is important to control vortex
mobility and consequently the dissipation produced by this motion.
Statistically, it is an example of a complex system consisting of elastic
objects immersed in a highly viscous media with long ranged interactions.
Quenched disorder, anisotropy and thermal fluctuations provide additional
elements that contribute to the richness of the subject \cite{nelson}.

The effect of an oriented potential in the dynamic response of the vortex
system is worth discussing. Obtaining the direction of motion it is a non
trivial task although certainly it would be affected by the presence of this
kind of potential. An extreme case would be a guided vortex motion where the
direction is determined by the subjacent symmetry \cite{staas,crabtree}.

In this letter we present results of electrical transport measurements on a
twin-oriented YBa$_{2}$Cu$_{3}$O$_{7}$ single crystal.  In order to test the
influence of twin plane boundaries on the vortex motion we have used a novel
technique with four pairs of coplanar contacts which allow us to control the
direction of the applied current. Two pairs were aligned as in a classical
transport experiment and two other similar pairs where oriented 90 degrees
with respect to the previous ones (see sketch in Fig.~1).  Using two current
sources through the external contacts we were able to rotate the current
direction in the {\em ab} plane in a continuous way. Measuring
simultaneously the voltages in both directions it is possible to extract
directly the velocity and {\em direction} of vortex motion. This permits us
to measure in the {\em same} crystal, the transport properties along and
across the twin planes and in any other orientation.

A crystal grown by the self flux technique \cite{gladys} having twin planes
oriented only in one direction was selected. Crystal dimensions where
$1\times1.1$ mm$^2$ in the {\em ab} plane and $75\,\mu$m in the {\em
c}-axis. Regularly spaced twins with an average separation of $5\,\mu$m
where observed in the crystal by microscope inspection under polarized
light.  Silver pads where sputtered on one of the faces through an {\em
ad-hoc} Cu mask and gold wires were attached with silver epoxy. DC
resistance measurements were performed as a function of temperature at
different magnetic fields applied in the {\em c}-axis using a two channel
nanovoltmeter.

\begin{figure}
\includegraphics[width=\columnwidth]{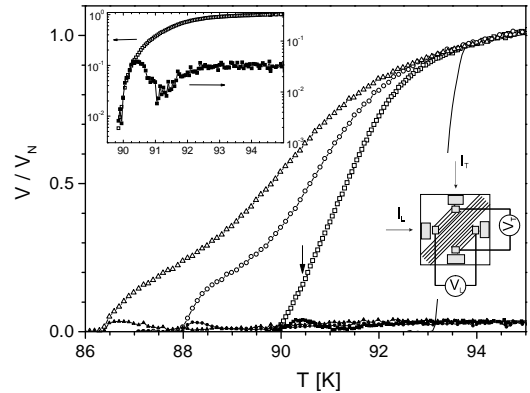} 
\caption{Resistive transitions at different applied fields for
an YBCO crystal, with unidirectional twin boundaries. The field is applied
in the {\em c}-direction and the current is applied in one of the  pairs
of current contacts. The open symbols correspond to the
voltage measured in the same direction as the applied current, the solid
ones to the voltage measured at the perpendicular direction.
Solid line: zero field; squares: 1T; circles: 2T; triangles: 3T.
\\Inset: Logarithmic plot of the measurement at 1T. The
perpendicular signal was scaled by a factor of 3
}
\label{fig1}
\end{figure}

In Fig.~\ref{fig1} we show the superconducting transitions for the longitudinal and
transverse voltage when the current is fed through one pair of the current
contacts. In this case the applied current makes an angle of 52 degrees with
the twin planes \cite{contactos} . The longitudinal component of the voltage shows the typical
broadening on magnetic field and close to the zero resistance temperature,
the kink usually associated with the onset of twin boundary pinning
\cite{flescher}. Close to this temperature superconducting coherence is
established along the whole sample thickness and in the field direction, as
has been reported through pseudo-dc-flux-transformer experiments
\cite{daniel}. Those experiments indicate that below certain temperature,
$T_{th}$, the vortices behave as continuous lines. As an example, the solid
arrow drawn in Fig.~\ref{fig1} marks the estimated $T_{th}$ for a magnetic
field of 1~T \cite{aclaro1}. At high temperatures the transversal component
of the voltage presents a small signal coming from a small misalignment of
the contacts and the Hall resistance \cite{hagen90}. Upon lowering the
temperature and close to the onset of twin boundary pinning, this component
starts to increase reaching, in our case, one third of the longitudinal
signal. This increase in the transverse voltage did not change on reversing
the magnetic field indicating that it is not related to the Hall resistance.

This is the first relevant result of the paper which indicates that the
vortex motion deviates strongly from the force direction at these
temperatures. The oblique motion can be attributed to vortex guidance by
twin planes \cite{crabtree}, or more generally by an anisotropic viscosity.
This latter results in an effective in-plane anisotropic resistivity tensor
with its principal axis (determined in this case by the twin planes
direction) rotated at an arbitrary angle respect to the current direction
\cite{aclaro}. In this situation it is well known that the electric field
and the current are not collinear \cite{landau}. More explicitly, the
longitudinal and transversal components of the voltage, neglecting the hall
resistance, are given by \begin{eqnarray} \label{eqV}
V_L=(\rho_\parallel\sin^2\alpha +\rho_\perp\cos^2\alpha )\, |{\bf j}_L|\\
\nonumber V_T=\left(\rho_\perp-\rho_\parallel\right)\sin\alpha\cos\alpha\,
|{\bf j}_L|, \end{eqnarray} where $\alpha$ is the angle between the current
and one of the principal axis of the system, $\rho_\perp$ and
$\rho_\parallel$ are the resistivity components across and along this
principal axis and ${\bf j}_L$ is the applied {\em longitudinal} current. It
is important to remark that this transverse voltage is {\em even} on
reversal of the magnetic field.

Shown in the inset of Fig.~\ref{fig1} are the same data for the case of an
applied field of 1T with the transverse signal scaled by a factor of 3. It
is clear from the data that for temperatures below 90.25 K, both signals are
proportional. From Eq.\ref{eqV} it can be seen that the ratio between
diagonal components of the resistivity tensor must be constant in this
temperature range. Furthermore, it strongly suggests that the transition to
the zero resistance state in both directions occurs at the same temperature.

Guided vortex motion implies that the velocity direction of vortices keeps
constant, independently of the external force. This can be tested by means
of the experiment described previously: using the extra pair of current
contacts, aligned at 90 degrees (therefore collinear with the transverse
voltage contacts) and by means of an additional current source, we rotate
continuously the current direction maintaining the net Lorentz force
constant (see sketch in Fig.~\ref{fig1}).

\begin{figure}
\includegraphics[width=\columnwidth]{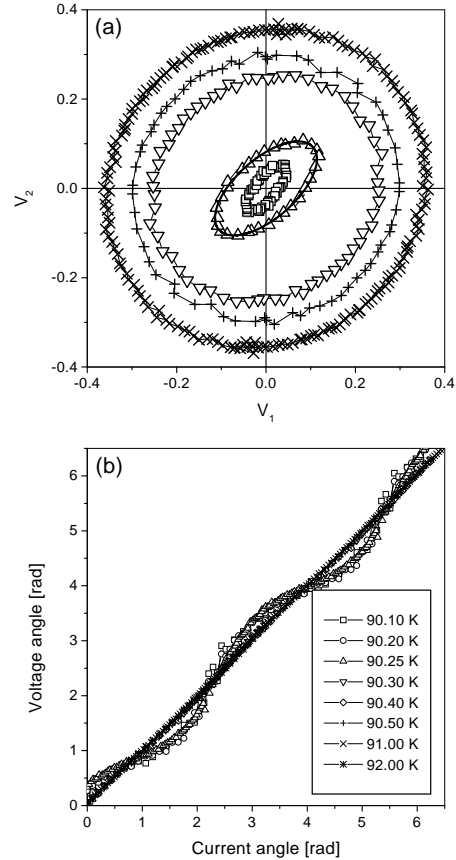} 
\caption{a) longitudinal vs. transversal voltage when varying the current
direction at a fixed current modulus and an applied field of 1T for
different temperatures as indicated in the graph. The solid line is a least
square fitting to the data at T = 90.25 K.  b)  Voltage angle 
vs. the current angle at different temperatures and an applied field of 1 T.
}
\label{fig2}
\end{figure}

In Fig.~\ref{fig2}(a) we plot the {\em longitudinal} component versus the
{\em transversal} component of the voltage when performing a current
rotation for different temperatures, at an applied field of 1T and for a
constant current density of $1.25\,\rm{A}/\rm{cm}^2$ (similar data where
obtained at 2~T and 3~T). As expected, for the high temperatures the data
lies along a circle reflecting the behavior of an isotropic conductor where
the value of the voltage response is independent of the current direction.
This also can be taken as a confidence test of current homogeneity in the
sample within the region of voltage contacts. On lowering the temperature
and approaching the kink the data starts to fall on an ellipse with its
principal axis oriented along the twin planes direction. The solid line
shown in the figure corresponds to a least square fit of the data at
$T=90.25$K using Eq.~\ref{eqV}, showing the excellent agreement with the
expected behavior of an anisotropic conductor. This elliptical response is
clearly observed in the whole liquid state with coherence along the c-axis
\cite{daniel}, without any evidence of a complete vortex guidance by twin
planes as was described before.

Shown in Fig.~\ref{fig2}(b) are the data plotted as electric field angle,
$\theta_E=\mbox{atan}(\frac{E_\parallel}{E_\perp})$, as a function of the
current density angle, $\theta_j=\mbox{atan}(\frac{j_\parallel}{j_\perp})$,
for different temperatures and with an applied magnetic field of 1T.  Again,
the data are representative of the evolution from isotropic to anisotropic
vortex motion. When being in the normal state as well as in the {\em
decoupled} vortex liquid phase the electric field direction strictly follows
the current direction. However, as the temperature is lowered and the
in-plane anisotropy develops, the electric field angle deviates from the
current angle.  
\begin{figure}
\includegraphics[height=4.5cm]{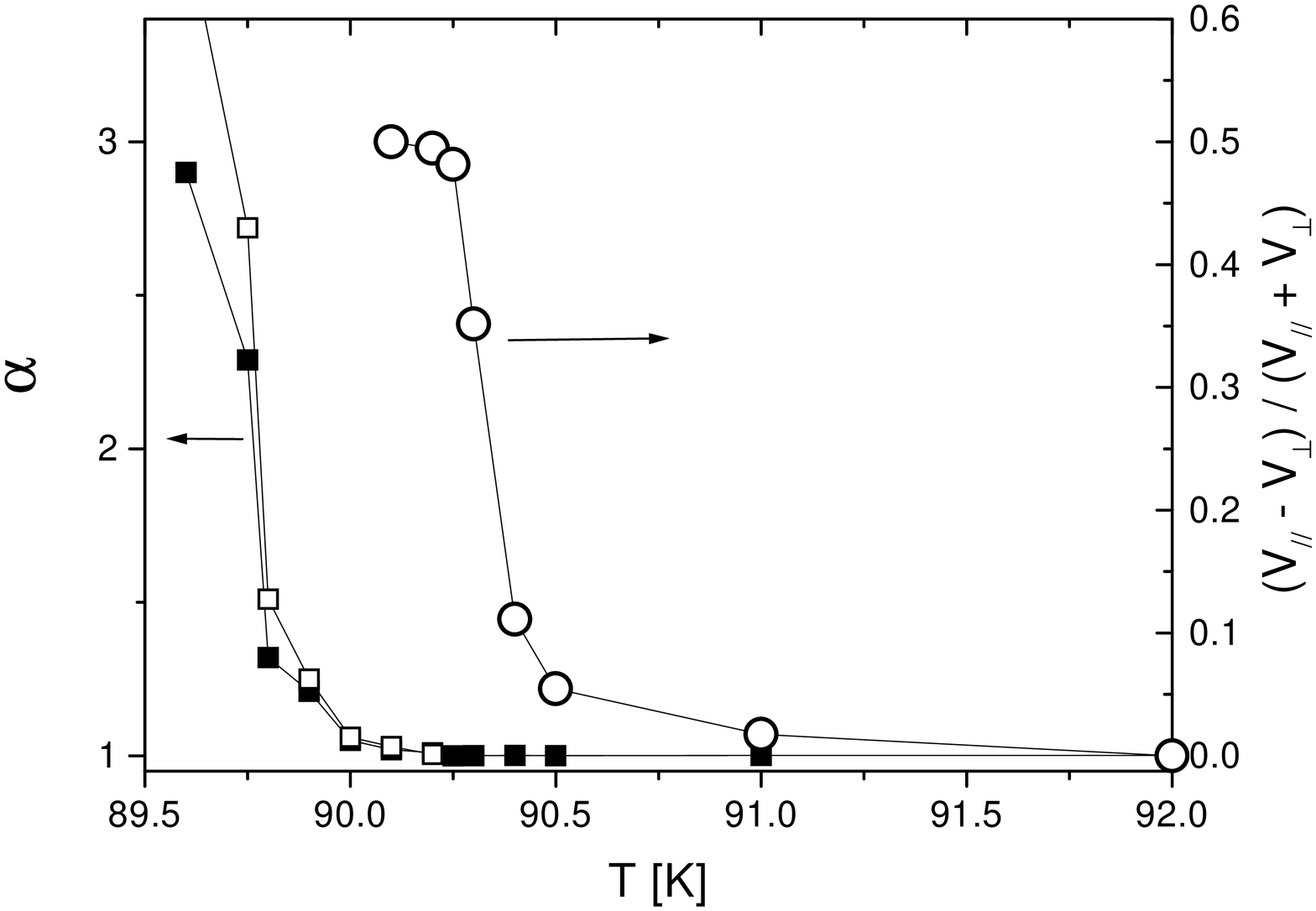}
\caption{Exponent taken from $V\simeq I^{\alpha}$ characteristics at
 different temperatures at an applied field of 1 T. Closed (Open) squares: 
perpendicular (parallel) component. Circles: Eccentricity functional
$\frac{V_{\parallel}-V_{\perp}}{V_{\parallel}+V_{\perp}}$ taken from minimum
square fits of the data. }
\label{fignew}
\end{figure}
In Fig.~\ref{fignew} the exponents,
$\alpha_{\parallel,\perp}$, are shown, taken from the I-V characteristics
for both directions and an applied field of 1 T, as a function of
temperature. Nonlinearities associated with the growth of the vortex solid
phase and the development of a critical current are observed below 89.9 K.
In the same graph we plot
$\frac{V_{\parallel}-V_{\perp}}{V_{\parallel}+V_{\perp}}$ obtained from
the current rotation experiment at different temperatures, which is as a
measure of the anisotropy in the resistivity.  It is evident from this
figure that this anisotropy in the vortex motion starts to develop in
a regime where the dynamics of vortex lines has a linear response.

Effectiveness of pinning by twin boundaries has been proved by low field
magnetic decorations \cite{pampa} and magneto-optical experiments
\cite{carlos,vlasko} showing that there is an increase of the vortex density
around twin boundaries. Although twin planes are relevant for the vortex
dynamics, the present data indicates that their main effect in the vortex
liquid state is to bring an anisotropic viscosity into its motion.
Resistance to vortex motion arises basically from two contributions:
Intrinsic viscosity (or friction) associated to the drag on a single vortex
(flux flow viscosity) and from vortex-vortex interaction. To explain the
observed anisotropic behavior by the single vortex friction we have to
include an anisotropy on the vortex characteristics, or take into account a
correction to the flux flow resistivity by pinning. The first argument can
be discarded because is very unlikely that an extrinsic conditional, like
the twin planes, can modify intrinsic superconducting properties starting
from certain temperature. The evaluation of the effect of an oriented
pinning landscape on vortex dynamics in the {\em linear} regime is a
difficult task. In the framework of thermally activated flux flow (TAFF)
\cite{kes} this effect is taken into account by changing the effective
number of moving vortices by thermal activation over the pinning potential.
The geometry of unidirectional pinning has been addressed by T. Matawari
\cite{mawatari} solving the Fokker-Planck equations which brings to an
anisotropy in the resistivity. However in this model, the ratio between
resistivities could be basically expressed as,
$\frac{\rho_\parallel}{\rho_\perp}=\exp(\frac{U_\perp-U_\parallel}{k_B T})$.
The observed temperature independence of this quantity in our data would
imply that twin planes pinning has a linear temperature 
dependence extrapolating to zero at zero temperature.

Other explanations trying to include modifications on the pinning landscape
by the external current like flux creep, necessarily end in a non-linear
behavior, in contradiction to the experimental data. In a recent paper D.
Lopez and coworkers \cite{daniel2} have observed a similar effect on
crystals with {\em splayed} columnar defects. In their work they explain the
experimentally observed anisotropy in the liquid taking into account a time
t$_{pl}$, during which the vortices resemble the solid phase. As a
consequence of the splayed defect configuration the solid phase has
anisotropic elastic constants that differentiates the responses in both
directions.

The reported results indicate an anisotropic dissipation below T$_{th}$. This
can be achieved if we suppose that at T$_{th}$, and below, pinning by the
twin planes becomes a center of nucleation of the entire vortex lines. At the
currents used in the experiments, these vortices remain strongly pinned in
the defects, generating a set of parallel irregular {\em grids} that the
vortex liquid has to cross when the force is oriented perpendicular to the
twin planes. Viscosity due to vortex entanglement is responsible for the
fact that the diffusion-like motion of vortices results different across and
along these {\em grids}. The density of vortices pinned in the twins only
depends on the density of these defects and does not depend either of
temperature or on current. This implies a saturated regime in which the
applied current is sufficiently below the critical current of the strongly
pinned vortices. Under these circumstances the relation between
resistivities would remain constant as a function of the temperature, as
observed.

\begin{figure}
\includegraphics[width=\columnwidth]{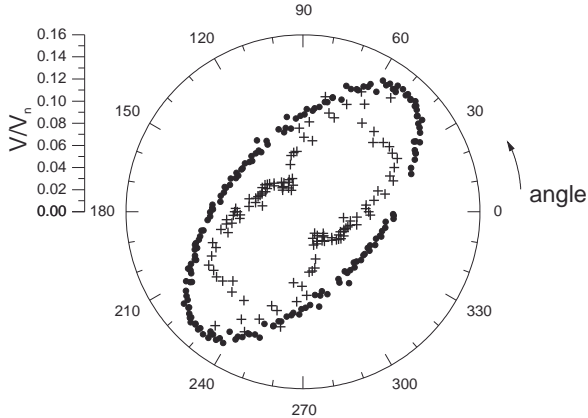} 
\caption{Polar graph of the voltage at different temperatures for an applied field of 1 T when rotating the current direction.
Circles: $T=90.25$ K and a current density of $1.25$ ${\rm A/cm^2}$. Crosses: $T=89.5$ K and a current density of $19$ ${\rm A/cm^2}$.
}
\label{fig3}
\end{figure}

Different and striking features are observed in the solid vortex phase. In
Fig.~\ref{fig3} we plot the data obtained in the rotating current experiment
for an applied field of 1 T at 89.5 K with a current density of
$19\,\rm{A}/\rm{cm}^2$, higher than the critical current density in both
directions. For comparison, the 90.25 K data taken with a current density
of $1.25\,\rm{A}/\rm{cm}^2$ is shown in the same graph. From this figure it
is clear that different mechanisms are involved in the vortex motion. The
elliptical shape evolves towards a squared eight-shaped curve, with a marked
minimum for the direction normal to the twin planes. Looking at the vortex
velocity across twins, the data shows a minimum in the motion when the force
is maximum along that direction. 
This component of velocity starts to
increase as soon as the velocity increases, reaching an almost constant
value.  Following the previous scenario, twin pinned vortices act like a
sieve for the others which are now condensed in a solid phase. To move
across twin planes vortices have to be able to explore laterally in
order to find weak channels to go through. In the liquid phase the lateral
displacement is occurring naturally because of the lack of shear modulus. On
the other hand, the same displacement in the solid, with a finite shear
stress capability \cite{pastoriza,kwok}, is only allowed when is
externally forced to move sideways.  Therefore, the experimental data
supports the existence of a guided motion along twin planes, with a constant
component of velocity in the perpendicular direction.

In summary, we have presented experimental data in a twin oriented crystal
that indicate an anisotropic vortex motion on a field dependent temperature
range. This anisotropy is different in the solid that in the liquid phase.
In the former there is evidence for a guided motion along the twins, whereas
in the liquid the response is consistent with an anisotropic vortex
viscosity.

We thank A.\ V.\ Silhanek for useful discussion and J. Luzuriaga for a
careful reading of the manuscript. This work was partially supported by the
Consejo Nacional de Investigaciones Cient\'{\i}ficas y Tecnol\'ogicas
(CONICET-PIP 4207) and the Agencia Nacional de Promoci\'on Cient\'{\i}fica y
Tecnol\'ogica (ANPCyT-PICT 00077 and 01116).


\begin{references}
\bibitem{contactos} The small rotation between contacts and the axis of the
crystal is a consequence of the manual alignment of the contact mask. The 
final orientation was estimated after the sputtering within 5 degrees and
checked with the fitting of the data in the rotating current experiment.
\bibitem{aclaro1} The estimation was done using the measured sample thickness and taking $T_{th}$ from ref.~\cite{daniel}
\bibitem{aclaro} The guided vortex motion can be taken as the limit of an extreme anisotropic viscosity, when the ratio between the diagonal components of the resistivity tensor goes to zero.
\end{references}
\end{document}